\documentclass[conference,letterpaper]{IEEEtran}

\usepackage{cite}
\usepackage{float}
\usepackage{multicol,multirow}
\usepackage{makecell,booktabs}
\usepackage{hyperref}
\hypersetup{    
    colorlinks=true,
    linkcolor=black,
    anchorcolor=black,
    citecolor=black,
    filecolor=black,
    menucolor=black,
    runcolor=black,
    urlcolor=black,
}

\usepackage{caption}
\usepackage{subcaption} 
\usepackage[utf8]{inputenc}

\usepackage{graphicx}
\usepackage{amsfonts}
\usepackage{amssymb}
\usepackage{amsmath,mathtools}
\usepackage{algorithmic}
\usepackage{algorithm}
\usepackage{relsize}
\usepackage{array}
\newcolumntype{P}[1]{>{\centering\arraybackslash}p{#1}}
\newcolumntype{L}[1]{>{\raggedright\arraybackslash}p{#1}}
\newcolumntype{R}[1]{>{\raggedleft\arraybackslash}p{#1}}
\usepackage{color}

\usepackage[labelsep=period]{caption}
\usepackage{bbm}
\usepackage{bm}
\usepackage{comment}
\usepackage{forest}
\usepackage{dblfloatfix}

\usepackage{etoolbox}

\title{\bf \Large
   Dispatch-Aware Deep Neural Network for Optimal Transmission Switching:\\Toward Real-Time and Feasibility Guaranteed Operation\vspace{-3mm}
}
\author{
    Minsoo Kim and Jip Kim\\
    Dept. of Energy Engineering,
    Korea Institute of Energy Technology\vspace{-3mm}
    \thanks{\vspace{-0mm}
    This work was supported by Basic Science Research Program through the National Research Foundation of Korea (NRF) funded by the Ministry of Education (No. RS-2023-00210018) and KENTECH Research Grant (202300008A).
    }    
    \vspace{-0mm}}

\begin{document}

\IEEEoverridecommandlockouts

\maketitle

\IEEEpubidadjcol

\begin{abstract}
    Optimal transmission switching (OTS) improves optimal power flow (OPF) by selectively opening transmission lines, but its mixed-integer formulation increases computational complexity, especially on large grids. To deal with this, we propose a dispatch‑aware deep neural network (DA‑DNN) that accelerates DC-OTS without relying on pre‑solved labels. DA‑DNN predicts line states and passes them through a differentiable DC‑OPF layer, using the resulting generation cost as the loss function so that all physical network constraints are enforced throughout training and inference. In addition, we adopt a customized weight-bias initialization that keeps every forward pass feasible from the first iteration, which allows stable learning on large grids. Once trained, the proposed DA‑DNN produces a provably feasible topology and dispatch pair in the same time as solving the DCOPF, whereas conventional mixed-integer solvers become intractable. As a result, the proposed method successfully captures the economic advantages of OTS while maintaining scalability.
\end{abstract}

\section*{Nomenclature}
\addcontentsline{toc}{section}{Nomenclature}
\begin{IEEEdescription}[\IEEEusemathlabelsep\IEEEsetlabelwidth{$\underline{\mathbf{p}}_g, \overline{\mathbf{p}}_g$}]
\item[$N_b$] Number of total buses in power network
\item[$N_g$] Number of generator buses in power network
\item[$N_l$] Number of lines in power network

\item[$\mathbf{p}_d, \mathbf{p}_g$] Real power demand / generation
\item[$\mathbf{\hat{p}}_g^*$] Predicted real power generation
\item[$\overline{\mathbf{p}}_g, \underline{\mathbf{p}}_g$] Maximum/minimum power generation limits
\item[$\Phi$] Line switching layers
\item[$\bm{\phi}$] Learnable parameters of $\Phi$
\item[$\mathcal{L}$] Loss function of $\Phi$
\item[$\mathbf{\hat{z}}$] Output of $\Phi$
\item[$\mathbf{\tilde{z}}$] Inputs of the last layer of $\Phi$
\item[${L}$] Lagrange function
\item[$\bm{\lambda}$] Lagrange multiplier of equality constraints
\item[$\bm{\mu}$] Lagrange multiplier of inequality constraints
\end{IEEEdescription}

\section{Introduction}

Optimal transmission switching (OTS) strategically opens or closes transmission lines so that the network topology becomes an active control variable \cite{fisher2008optimal}. By doing so, system operators can reroute power flows and reduce generation costs. This effect is known as Braess’s paradox in power systems, that removing a line can redirect flows and lower the total operational cost \cite{schafer2022understanding}. Large‑scale simulation-based analysis shows that the payoff is tangible. For example, simulations from the ARPA-E Topology Control Algorithm project show that applying switching actions on the PJM system can cut congestion costs by more than 50\% and save over \$100 million annually in the real-time market \cite{ARPA_E2014}.

Reconizing both its benefits and computational challenges, regulators now require that tranmission switching be evaluated alongside other grid‑enhancing technologies. For example, FERC Order No. 1920 in the United States and ENTSO‑E’s R\&I Roadmap 2017‑2026 in Europe list transmission switching as an option worth considering \cite{fercorder1920, entsoe2017roadmap}. Furthermore, system operators such as PJM and ISO‑NE already apply limited corrective switching during contingencies \cite{PJMManual37, ISONEOP19}.

Despite its potential, the OTS problem is computationally intractable. Even in DC formulations (DC-OTS), the presence of binary line-status variables converts the task into a mixed integer optimization that is NP-hard \cite{kocuk2016cycle}. Unfortunately, commercial solvers can handle only moderate network sizes when solving OTS. For example, as the number of candidate lines grows to hundreds or thousands, the solution times escalate well beyond the operational limits, often exceeding an hour for realistic cases \cite{hedman2008optimal}. Hence, the economic and reliability benefits of OTS cannot be fully realized without first alleviating its computational burden. This has motivated a stream of research on faster solution techniques \cite{yang2019line}.

\begin{figure*}[t]
	\centering
\includegraphics[width=0.9\textwidth]{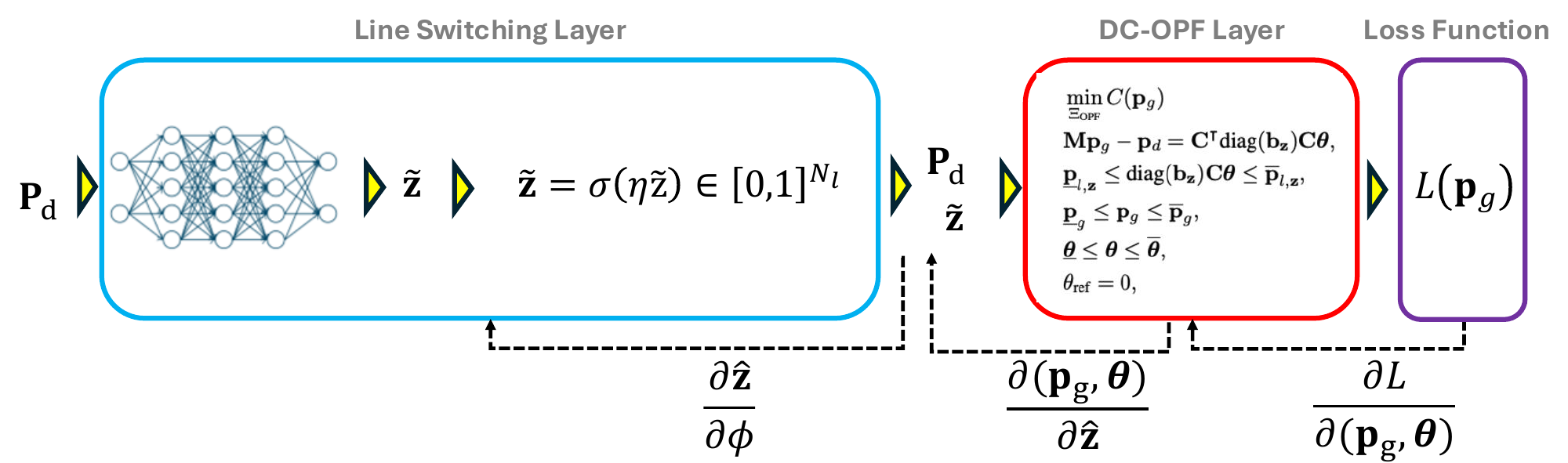}
	\caption{Training process of the proposed dispatch-aware deep neural network for optimal transmission switching.}
	\label{fig:DA-DNN_framework}\vspace{-3mm}
\end{figure*}

Researchers have recently turned to learning-based approaches to overcome the computational barrier of OTS. These data-driven techniques leverage historical optimal solutions or simulation data to train models that guide, warm-start, or even replace parts of the optimization. In \cite{yang2019line}, machine learning (ML) was used to rank and prioritize the candidate of line-switching status. In doing so, they effectively narrow the search space that the solver must explore. In \cite{pineda2024learning}, supervised classifiers, such as k-nearest neighbor (kNN), have also been trained on the optimal grid topology to identify which lines should remain open or closed under certain conditions. The more advanced ML technique was adopted in \cite{bugaje2023real}. Specifically, the authors of \cite{bugaje2023real} used a radial-basis function neural network (RBFNN) that directly outputs which lines to open given system state. Although these works significantly accelerate the OTS solving process, their methods are based on supervised learning. Since solving OTS is computationally expensive, obtaining presolved solutions as labels takes an enormous amount of computation time. \cite{tang2022optimal} circumvented this challenge by employing deep reinforcement learning to learn switching policies without presolved labels; however, the feasibility of the resulting decisions is not guaranteed. 

To overcome these two challenges, i.e., the difficulty of obtaining presolved OTS solutions as training labels and the need to guarantee feasibility, we propose a novel dispatch-aware learning framework for solving DC-OTS. To deal with the first challenge, the proposed method uses line switching layer that predicts line status values and DC-OPF layer solves DC-OPF with predicted line status to train the model by directly minimizing the generation cost without using any labels. For the second challenge, the embedded DC-OPF layer enforces all the constraints during every forward pass, and obtains feasibility guaranteed dispatch from the DNN.

We summarize our key contributions as follows:

\begin{enumerate}
    \item We propose a novel dispatch-aware deep neural network (DA-DNN) to accelerate solving DC-OTS which consists of a line switching layer and a DC-OPF layer. During training, the line switching layer outputs predicted line statuses. Next, the embedded DC-OPF layer solves DC-OPF with these line statuses to obtain the generator dispatch. The resulting generation cost is used as an unsupervised loss, which eliminates the need for precomputed OTS as labels. Since every forward pass solves a full DC-OPF, all physical constraints are enforced throughout training and inference. After training, we simply binarize the relaxed line status and run DC-OPF to obtain a pair of line status and dispatch that is provably feasible. As a result, the total computation time for our method in the inference phase is always equal to that of a single DC-OPF. To the best of our knowledge, this is the first work that solve DC-OTS using unsupervised learning while preserving all the physical network constraints.

    \item We introduce a weight and bias initialization scheme that ensures that the embedded DC-OPF is feasible from the first iteration. This enables reliable training on large systems, whereas without this initialization the framework fails training, since the DC-OPF repeatedly returns infeasible solutions.
    
    \item We evaluate the proposed framework on the IEEE 73- and 300-bus systems. The results show that each inference finishes in the same computation time as a single DC-OPF solve. Moreover, a commercial solver fails to complete DC-OTS in the 300-bus case within an hour, whereas the proposed framework produces line statuses that lower generation cost relative to DC-OPF in just milliseconds.
    
\end{enumerate}

\section{DC Optimal Transmission Switching Model}

Suppose that the given power system consists of $N_b$ buses, $N_g$ generators, and $N_l$ lines. Let $\Xi_\text{OTS}:=\{\mathbf{p}_g, \bm{\theta}, \mathbf{z}\}$ be the set of optimization variables, where $\mathbf{p}_g\in\mathbb{R}^{N_g}$, $\bm{\theta}\in\mathbb{R}^{N_b}$, and $\mathbf{z}\in\{0,1\}^{N_l}$ denote active power generation, voltage phase angle, and line on/off status. Then, the DC-OTS problem is formulated as follows \cite{ahmadi2023decomposition}:
\begingroup
\allowdisplaybreaks
\begin{subequations}\label{Eq:DC-OTS}
\begin{align}
&
\min_{\Xi_\text{OTS}}
C(\mathbf{p}_g)
\label{Eq:OTS_Obj}\\
&
\mathbf{M}\mathbf{p}_g - \mathbf{p}_d = \mathbf{C}^\intercal \big(\mathbf{z}\odot\text{diag}(\mathbf{b}_l)\mathbf{C}\bm{\theta}\big),\label{Eq:OTS_PF}\\
&
\mathbf{z}\odot\underline{\mathbf{p}}_l \leq \mathbf{z}\odot\text{diag}(\mathbf{b})\mathbf{C}\bm{\theta} \leq \mathbf{z}\odot\overline{\mathbf{p}}_l,\label{Eq:OTS_line_limits}\\
&
\underline{\mathbf{p}}_g \leq \mathbf{p}_g \leq \overline{\mathbf{p}}_g,\label{Eq:OTS_gen_limits}\\
&
\underline{\bm{\theta}}\leq \bm{\theta} \leq \overline{\bm{\theta}},\label{Eq:OTS_angle_limits}\\
&
\theta_\text{ref} = 0, \label{Eq:OTS_slack}
\end{align}
\end{subequations}
\endgroup
\noindent 
where $C(\cdot):\mathbb{R}^{N_g}\rightarrow\mathbb{R}$ in (\ref{Eq:OTS_Obj}) is the convex generation cost function of each generator. (\ref{Eq:OTS_PF}) enforces DC power balance equation, where $M\in \{0,1\}^{N_b \times N_g}$ is generator-bus incidence matrix that maps $\mathbf{p}_g$ to the vectors of $N_b$ dimensions, $\mathbf{p}_d\in\mathbb{R}^{N_b}$ denotes nodal demand of each bus, and $\mathbf{C}^\intercal\big(\mathbf{z}\odot\text{diag}(\mathbf{b}_l)\mathbf{C}\bm{\theta}\big)$ represents the line flows. Here, $\mathbf{C}\in\{-1,0,1\}^{N_l\times N_b}$ is the branch-bus incidence matrix, $\mathbf{b}_l\in\mathbb{R}^{N_l}$ is the vector of line suceptances, and $\odot$ refers to element-wise multiplication. Note that $\mathbf{z}$ nullifies any line flows that are switched out. These line flows are restricted by $\underline{\mathbf{p}}_l$ and $\overline{\mathbf{p}}_l$ in (\ref{Eq:OTS_line_limits}). In addition, the output of the generator and the angle of the nodal voltage are limited by (\ref{Eq:OTS_gen_limits}) and (\ref{Eq:OTS_angle_limits}), respectively. Finally, (\ref{Eq:OTS_slack}) defines the slack bus.

\section{Methodologies}

Unlike DC-OPF, which can be solved in milliseconds, DC-OTS is NP-hard due to the binary vector $\mathbf{z}$. Thus, commercial MILP solvers may require several hours or even days to reach optimality. Consequently, practical adoption hinges on accelerated solution techniques. Thus, in this section, we propose a dispatch-aware deep neural network (DA-DNN), an efficient DNN-based method that obtains near-optimal switching decisions within operational time frames.

The overall framework of the proposed DA-DNN is illustrated in Fig.~\ref{fig:DA-DNN_framework}. As shown in the figure, DA-DNN consists of a line switching layer and a DC-OPF layer. The line switching layer takes bus demand and predicts the line status of the network. Then, DC-OPF layer takes the predicted line status, solves DC-OPF based on that values, and obtains dispatch. After that, the model learns to reduce the generation cost that is determined from the obtained dispatch. Note that we don't need to prepare any presolve labes of DC-OTS when training the proposed DA-DNN.

\subsection{Forward Pass of DA-DNN}

\subsubsection{Solving DC-OPF with NN predicted line status}
Let $\Phi:\mathbb{R}^{N_b}\rightarrow\mathbb{R}^{N_l}$ be the line switching layer in Fig.~\ref{fig:DA-DNN_framework} that maps $\mathbf{p}_d$ to the predicted line switching state $\mathbf{\hat{z}}\in [0, 1]^{N_l}$. Note that all elements of $\mathbf{\hat{z}}$ are bounded by $[0,1]$, since $\mathbf{\hat{z}}$ is used as a continuous relaxation of the binary line switch state when solving (\ref{Eq:DC-OTS_relax}). For this bound, we use a sigmoid function $\sigma(\cdot)$ as follows:
\begin{equation}
    \mathbf{\hat{z}} = \sigma\Big(\eta(\mathbf{W}\mathbf{\tilde{z}} + \mathbf{b})\Big)
\end{equation}
where $\mathbf{W}\in\mathbb{R}^{N_h \times N_l}$, $\mathbf{b}\in\mathbb{R}^{N_l}$, and $\mathbf{\tilde{z}}$ are the weight matrix, the bias vector, and the input of the last layer of $\Phi$, respectively. In addition, $\eta\in\mathbb{R}$ is the scaling factor and $\eta\geq 1$. The purpose of using the scaling factor $\eta$ is to reduce the binarization error of $\mathbf{\hat{z}}$.

Now, we solve the DC-OPF using $\mathbf{\hat{z}}$. Let $\Xi_\text{OPF} := \{\mathbf{p}_g, \bm{\theta} \}$ be the set of optimization variables of DC-OPF. Then, the optimization problem is formulated as follows:
\begingroup
\allowdisplaybreaks
\begin{subequations}\label{Eq:DC-OTS_relax}
\begin{align}
&
\min_{\Xi_\text{OPF}}
C(\mathbf{p}_g)
\label{Eq:OTS_z_Obj}\\
&
\mathbf{M}\mathbf{p}_g - \mathbf{p}_d = \mathbf{C}^\intercal \text{diag}(\mathbf{b}_\mathbf{z})\mathbf{C}\bm{\theta},\label{Eq:OTS_z_PF}\\
&
\underline{\mathbf{p}}_{l,\mathbf{z}} \leq\text{diag}(\mathbf{b}_{l,\mathbf{z}})\mathbf{C}\bm{\theta} \leq \overline{\mathbf{p}}_{l,\mathbf{z}},\label{Eq:OTS_z_line_limits}\\
&
\text{(\ref{Eq:OTS_gen_limits})--(\ref{Eq:OTS_slack})},\label{Eq:OPF_same_constraints}
\end{align}
\end{subequations}
\endgroup
\noindent
where $\mathbf{b}_{l,\mathbf{z}} = \mathbf{z}\odot\mathbf{b}_l$, $\underline{\mathbf{p}}_{l,\mathbf{z}} = \mathbf{z}\odot\underline{\mathbf{p}}_{l}$, and $\overline{\mathbf{p}}_{l,\mathbf{z}} = \mathbf{z}\odot\overline{\mathbf{p}}_{l}$. This optimization problem is DC-OPF with a continuously relaxed line switching state $\mathbf{\hat{z}}$. However, in the initial training process, (\ref{Eq:DC-OTS_relax}) is likely infeasible. Thus, stable training of the proposed DA-DNN is impossible. 

Let \(\mathbf W_{\mathrm{init}}\) and \(\mathbf b_{\mathrm{init}}\) denote the initial weight matrix and bias vector of the last layer of \(\Phi\), respectively.  
The resulting initial relaxed line state $\mathbf{\hat{z}}_\text{init}$ is
\begin{equation}
    \hat{\mathbf z}_{\mathrm{init}}
  =\sigma\!\Bigl(\eta\bigl(\mathbf W_{\mathrm{init}}\tilde{\mathbf z}_{\mathrm{init}}
                        +\mathbf b_{\mathrm{init}}\bigr)\Bigr),
\end{equation}
where $\tilde{\mathbf z}_{\mathrm{init}}$ is the input to the last initialized layer.  
We set \(\mathbf W_{\mathrm{init}}=\mathbf 0\) and \(\mathbf b_{\mathrm{init}}=9/\eta\), so that every component is evaluated at $\mathbf{\hat{z}}_\text{init} = \sigma(9) = 0.9999$. Employing $\hat{\mathbf z}_{\mathrm{init}}$ for~\eqref{Eq:DC-OTS_relax} solves the optimization problem with the standard DC-OPF. 

\subsubsection{Loss function to train the model} Let $\Xi_\text{OPF}^* := \{\mathbf{\hat{p}}_g^*, \bm{\hat{\theta}}^*\}$ be the set of solutions of (\ref{Eq:DC-OTS_relax}), which is defined as follows:
\begin{equation}
    \Xi_\text{OPF}^* := \{\mathbf{\hat{p}}_g^*, \bm{\hat{\theta}}^*\} = \arg\min_{\Xi_\text{OPF}\in\mathcal{F}(\mathbf{\hat{z}})}C(\mathbf{p}_g)
    \label{Eq:argmin_function}
\end{equation}
where $\mathcal{F}(\mathbf{\hat{z}})$ is the feasible region determined from $\mathbf{\hat{z}}$. Now, we determine the loss function using $\Xi_\text{OPF}^*$ to train $\Phi$ as $\mathcal{L}(\mathbf{\hat{p}}_g)~=~C(\mathbf{\hat{p}}_g)$. Thus, we directly minimize the overall generation cost in an unsupervised learning framework, i.e., we do not need to prepare the presolved solution as labels. In the inference phase, we binarize $\mathbf{\hat{z}}$ based on the threshold (e.g., 0.5), since real-world transmission switching status has only on/off status.

\subsection{Backpropagation Through DC-OPF Layer}

The backpropagation to train $\Phi$ with a learnable parameter $\bm{\phi}$ (e.g., $\mathbf{W}$ and $\mathbf{b}$) is expressed as follows:
\begin{equation}
    \nabla_{\bm{\phi}} \mathcal{L} = \frac{\partial \mathcal{L}}{\partial (\mathbf{\hat{p}}_g^*, \mathbf{\hat{\theta}}^*)}\frac{\partial (\mathbf{\hat{p}}_g^*, \mathbf{\hat{\theta}}^*)}{\partial \mathbf{\hat{z}}}\frac{\partial \mathbf{\hat{z}}}{\partial \bm{\phi}}.\label{Eq:dd}
\end{equation}
Here,  $\frac{\partial \mathcal{L}}{\partial (\mathbf{\hat{p}}_g^*, \mathbf{\hat{\theta}}^*)}$ is determined analytically since $\mathcal{L}(\mathbf{\hat{p}}_g)=C(\mathbf{\hat{p}}_g)$.  $\frac{\partial \mathbf{\hat{z}}}{\partial \bm{\phi}}$ is determined from automatic differentiation, which is widely adopted to train deep learning models. So, we need to obtain $\frac{\partial (\mathbf{\hat{p}}_g^*, \mathbf{\hat{\theta}}^*)}{\partial \mathbf{\hat{z}}}$, which requires derivatives of the argmin function (\ref{Eq:argmin_function}).

For this, we simplify the DC-OPF problem (\ref{Eq:DC-OTS_relax}) as follows:
\begingroup
\allowdisplaybreaks
\begin{subequations}\label{Eq:simple_OPF}
\begin{align}
&
\min_{\mathbf{x}}
f(\mathbf{x})
\label{Eq:simple_obj}\\
&
g(\mathbf{x},  \mathbf{\hat{z}}) = 0,\label{Eq:simple_PF}\\
&
h(\mathbf{x},  \mathbf{\hat{z}}) \leq 0,\label{Eq:simple_limits}
\end{align}
\end{subequations}
\endgroup
\noindent
where $\mathbf{x}$ denotes the optimization variables of (\ref{Eq:DC-OTS_relax}). Then, the Lagrange function $L$ of the optimization problem is formulated as:
\begin{equation}
    L(\mathbf{x}, \bm{\lambda}, \bm{\mu}; \mathbf{\hat{z}}) = f(\mathbf{x}) + \bm{\lambda}^\intercal g(\mathbf{x},  \mathbf{\hat{z}}) + \bm{\mu}^\intercal h(\mathbf{x},  \mathbf{\hat{z}})
\end{equation}
where $\bm{\lambda}$ and $\bm{\mu}$ are the dual variables of equality and inequality constraints, respectively. Since (\ref{Eq:simple_OPF}) is a convex optimization problem, the KKT conditions are sufficient and necessary conditions for optimality. Thus, we have
\begin{equation}
    \mathcal{I} := \begin{bmatrix}
        \frac{\partial L(\mathbf{x}, \bm{\lambda}, \bm{\mu}; \mathbf{\hat{z}})}{\partial \mathbf{x}}\\
        g(\mathbf{x}, \mathbf{\hat{z}})\\
        h(\mathbf{x}, \mathbf{\hat{z}})
        \end{bmatrix}  =  \mathbf{0}
\end{equation}
where $\mathcal{I}$ is the implicit function of $\mathbf{x}$ and $\mathbf{\hat{z}}$ for the KKT optimality conditions. Since the derivative of $\mathcal{I}$ with respect to $\mathbf{\hat{z}}$ exists due to the implicit function theorem \cite{krantz2002implicit}, we have
\begin{equation}
    \frac{\partial \mathcal{I}}{\partial \mathbf{\hat{z}}} + \frac{\partial \mathcal{I}}{\partial \mathbf{x}}\frac{\partial \mathbf{x}}{\partial \mathbf{\hat{z}}} = \mathbf{0}.
\end{equation}
Since $\mathbf{x}$ represents the optimization variables $\mathbf{\hat{p}}_g^*$ and $ \mathbf{\hat{\theta}}^*$, we have
\begin{equation}
    \frac{\partial (\mathbf{\hat{p}}_g^*, \mathbf{\hat{\theta}}^*)}{\partial \mathbf{\hat{z}}} = -\frac{\partial \mathcal{I}}{\partial \mathbf{\hat{z}}} \bigg(\frac{\partial \mathcal{I}}{\partial (\mathbf{\hat{p}}_g^*, \mathbf{\hat{\theta}}^*)}\bigg)^{-1}.
\end{equation}
It should be noted that $\frac{\partial \mathcal{I}}{\partial \mathbf{\hat{z}}}$ and $\frac{\partial \mathcal{I}}{\partial \mathbf{x}}$ are both explicitly determined. Hence, we now have the backpropagation process to update $\bm{\phi}$ of $\Phi$ as follows:
\begin{equation}
    \nabla_{\bm{\phi}} \mathcal{L} = -\frac{\partial \mathcal{L}}{\partial (\mathbf{\hat{p}}_g^*, \mathbf{\hat{\theta}}^*)}\frac{\partial \mathcal{I}}{\partial \mathbf{\hat{z}}} \bigg(\frac{\partial \mathcal{I}}{\partial (\mathbf{\hat{p}}_g^*, \mathbf{\hat{\theta}}^*)}\bigg)^{-1}\frac{\partial \mathbf{\hat{z}}}{\partial \bm{\phi}}.\label{Eq:dd}
\end{equation}
Using this result, the learnable parameter $\bm{\phi}$ is updated as follows:
\begin{equation}
    \bm{\phi} \leftarrow \bm{\phi} + \eta_\text{lr}\frac{\partial \mathcal{L}}{\partial (\mathbf{\hat{p}}_g^*, \mathbf{\hat{\theta}}^*)}\frac{\partial \mathcal{I}}{\partial \mathbf{\hat{z}}} \bigg(\frac{\partial \mathcal{I}}{\partial (\mathbf{\hat{p}}_g^*, \mathbf{\hat{\theta}}^*)}\bigg)^{-1}\frac{\partial \mathbf{\hat{z}}}{\partial \bm{\phi}}
\end{equation}
where $\eta_\text{lr}$ denotes the learning rate of DA-DNN.

\section{Case Study}
\subsection{Simulation Settings}

We use two PGLib-OPF networks, i.e., IEEE 73 bus system and IEEE 300 bus system, to evaluate DA-DNN \cite{babaeinejadsarookolaee2019power}. We generate 3,000 samples from the uniform distribution of 100$-$110\% loading level. Note that we discard the samples that are infeasible for DC-OPF. We divide the samples into 3:1:2 for training, validation, and test.

We use three layers of multilayer perceptron. The size of the hidden vectors is 64 for IEEE 73 bus system and 128 for IEEE 300 bus system. We employ the exponential linear units (ELU) activation function at the outputs of each layer, except the last layer of the line switching layer \cite{clevert2015fast}.We use AdamW optimizer with a learning rate of $5\times10^{-5}$ and a weight decay of $10^{-2}$ for training \cite{loshchilov2017decoupled}. The maximum training epochs are 50 with a mini-batch size of 25. We use CvxpyLayer \cite{agrawal2019differentiable} to formulate DC-OPF and ECOS to solve the optimization problem during training. For testing, we use the Gurobi solver for all the optimization based approaches \cite{gurobi}.

\subsection{Overall Performance Comparisions}

\begin{table}[t]
	\centering
	\caption{Results in IEEE 73 bus system.}
	\label{table:73bus_result}
	\begin{tabular}{c|cc|cc}
		\toprule
		\multirow{3.5}{*}
        {\makecell{Method}}&\multicolumn{2}{c|}{$\overline{\bm{\theta}} = 0.35$}&\multicolumn{2}{c}{$\overline{\bm{\theta}} = 0.4$}\\ 
        \cmidrule(lr){2-5}
        & \makecell{Avg. Gen.\\Cost (\$1k)}  &\makecell{Avg. Comp.\\Time (sec)} &\makecell{Min. Gen.\\Cost (\$1k)} &\makecell{Avg. Comp.\\Time (sec)}\\\midrule\midrule
	    \makecell{ED\\(Gurobi)} & \makecell{183.01} & 0.00 & 183.01&0.00\\
		\makecell{DC-OPF\\(Gurobi)} & \makecell{183.24} & 0.00 & 183.02 &0.00\\
		\makecell{DC-OTS\\(Gurobi)} & 183.04 & 5.23 & 183.01 & 3.54\\
        \midrule
		\makecell{DC-OTS\\(\textbf{DA-DNN})} & \makecell{183.10} & \makecell{0.00} & \makecell{183.01} &0.00\\
		\bottomrule
	\end{tabular}
    \captionsetup{justification=raggedright, singlelinecheck=false}
\end{table}

\begin{table}[t]
	\centering
	\caption{Results in IEEE 300 bus system.}
	\label{table:300bus_result}
	\begin{tabular}{c|cc|cc}
		\toprule
		\multirow{3.5}{*}
        {\makecell{Method}}&\multicolumn{2}{c|}{$\overline{\bm{\theta}} = 0.5$}&\multicolumn{2}{c}{$\overline{\bm{\theta}} = 0.6$}\\ 
        \cmidrule(lr){2-5}
        & \makecell{Avg. Gen.\\Cost (\$1k)}  &\makecell{Avg. Comp.\\Time (sec)} &\makecell{Avg. Gen.\\Cost (\$1k)} &\makecell{Avg. Comp.\\Time (sec)}\\\midrule\midrule
	    \makecell{ED\\(Gurobi)} & \makecell{479.81} & 0.00& 479.81&0.00\\
		\makecell{DC-OPF\\(Gurobi)} & \makecell{528.98} & 0.00& 523.04&0.00\\
		\makecell{DC-OTS\\(Gurobi)} & NS. & NS. & NS. & NS.\\
        \midrule
		\makecell{DC-OTS\\(\textbf{DA-DNN})} & \makecell{520.77} & \makecell{0.00} & \makecell{514.34} &0.00\\
		\bottomrule
	\end{tabular}
    \captionsetup{justification=raggedright, singlelinecheck=false}
  \caption*{\quad NS.: not solved within an hour.}
\end{table}

Table~\ref{table:73bus_result} and Table~\ref{table:300bus_result} compare four operational strategies on the IEEE 73‑ and 300‑bus test systems. Economic dispatch (ED) provides the economic lower bound because it ignores all network constraints, whereas DC‑OPF with every line in service constitutes an upper bound on achievable savings. DC‑OTS denotes the mixed‑integer benchmark solved with Gurobi. The proposed DA‑DNN predicts one set of relaxed line‑status values and binarized based on 0.5. After that, a subsequent DC‑OPF is then solved to obtain the final dispatch and operating cost.

In the IEEE 73 bus system, topology control is valuable when the bus voltage angle is relatively tight ($\overline{\bm{\theta}} = 0.35$). In this case, the proposed DA‑DNN lowers the generation cost of DC‑OPF from \$183.24k to \$183.10k, which is roughly 60\% of the maximum generation cost relief obtainable between the OPF and ED while achieving the same computational time as a DC‑OPF (milliseconds). Although DC-OTS attains a marginally lower cost of \$183.04k, it spends 5.23 sec per snapshot, which is two orders of magnitude slower than the proposed DA-DNN. When the angle limit is relaxed to $\overline{\bm{\theta}} = 0.4$, most line flow bottlenecks disappear, so all methods converge to nearly the same cost; interestingly, unlike DC-OTS with solvers, DA‑DNN neither degrades performance in generation cost nor loses its advantages in computation time.

The scalabiity benefit of the DA-DNN is clearer on  the IEEE 300 bus system. With $\overline{\bm{\theta}} = 0.5$, DC-OPF incurs \$528.98k, whereas DA-DNN obtains a feasible topology with 1.6\% reduced generation cost. Note that DC-OTS with a commercial solver is unable to produce an optimal solution within one hour, confirming that exhaustive switching becomes intractable at this scale. Similar behavior is observed at the looser angle limit of $\overline{\bm{\theta}} = 0.6$, where DA-DNN still lowers cost by 1.7\% from the DC-OPF.

Since the DC-OPF layer is embedded in every forward pass, all DA-DNN topologies automatically satisfy the physical constraints. This benefit makes the outputs directly deployable for real-time operation. Thus, the proposed method combines the speed of DC-OPF with the economic benefits of full OTS.

\subsection{Impact of the weight and-bias initialization}

\begin{figure}[t]
	\centering
\includegraphics[width=1\columnwidth]{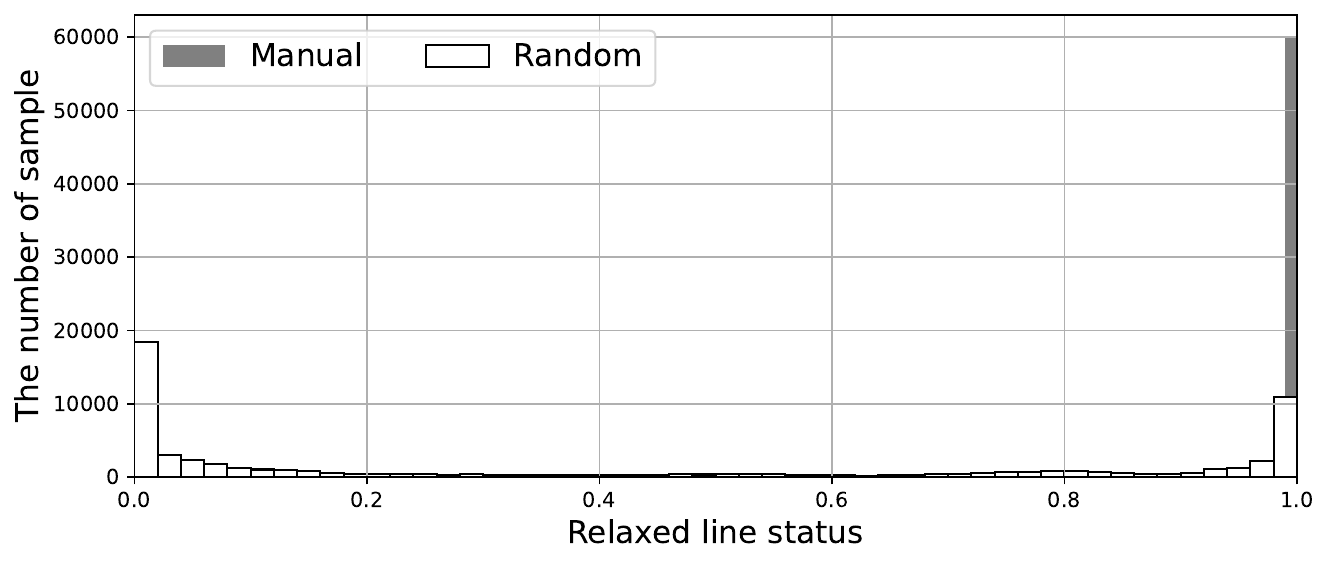}
	\caption{Histogram of the predicted relaxed line status values from untrained DA-DNN with different weight and bias initialization.}
	\label{fig:weight_histogram}
\end{figure}

\begin{figure}[t]
     \centering
     \begin{subfigure}[b]{0.49\columnwidth}
     \centering
         \includegraphics[width=\columnwidth]{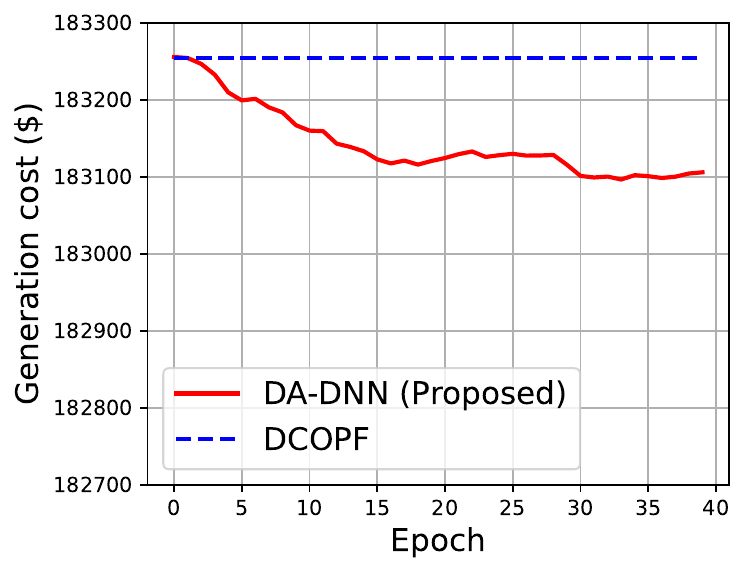}
         \caption{\small $\overline{\bm{\theta}} = 0.35$.}
         \label{fig:theta_0p4_case73}
     \end{subfigure}
     \begin{subfigure}[b]{0.49\columnwidth}
     \centering
         \includegraphics[width=\columnwidth]{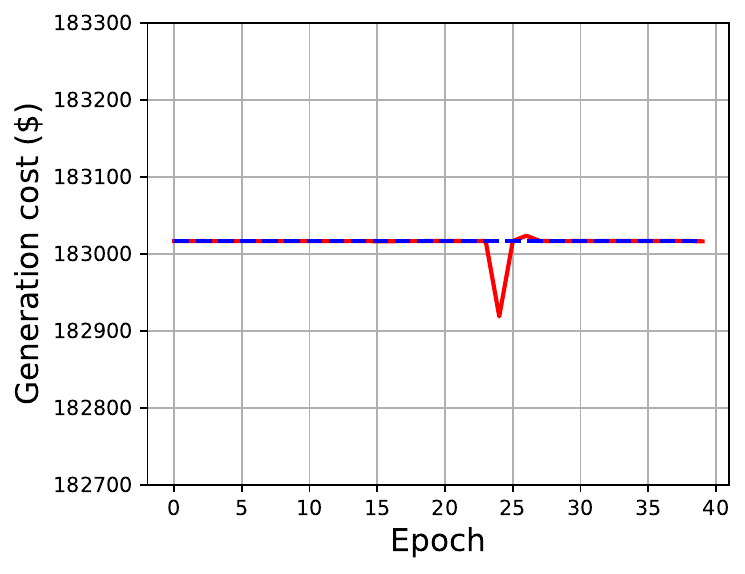}
         \caption{\small $\overline{\bm{\theta}} = 0.4$.}
         \label{fig:theta_0p35_case73}
     \end{subfigure}
     \caption{\small Training loss curve in IEEE 73 bus system.}
     \label{fig:result_73}
\end{figure}

\begin{figure}[t]
     \centering
     \begin{subfigure}[b]{0.48\columnwidth}
     \centering
         \includegraphics[width=\columnwidth]{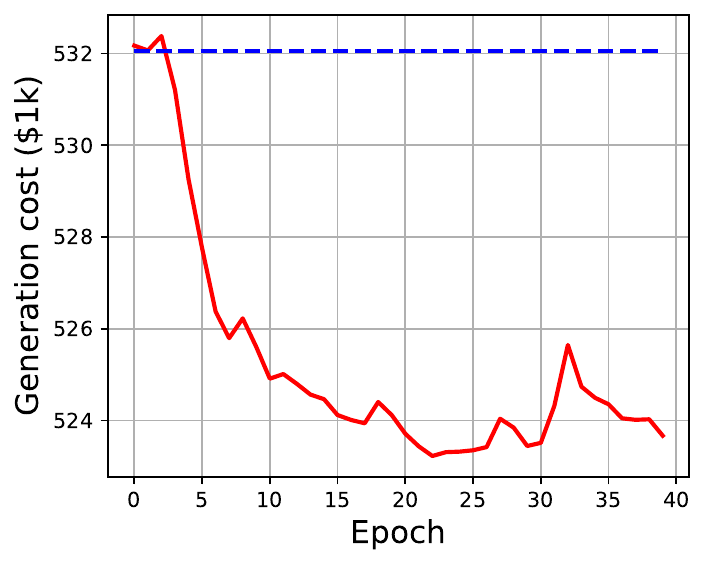}
         \caption{\small $\overline{\bm{\theta}} = 0.5$.}
         \label{fig:theta_0p5_case300}
     \end{subfigure}
     \begin{subfigure}[b]{0.48\columnwidth}
     \centering
         \includegraphics[width=\columnwidth]{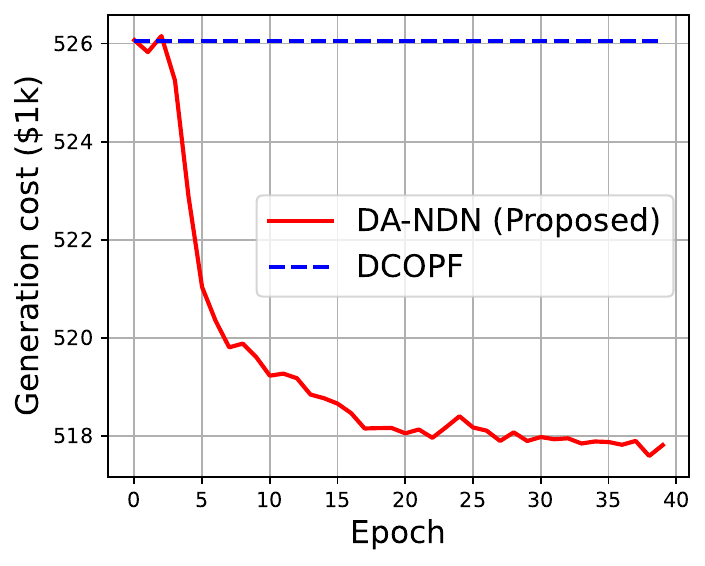}
         \caption{\small $\overline{\bm{\theta}} = 0.6$.}
         \label{fig:theta_0p6_case300}
     \end{subfigure}
     \caption{\small Training loss curve in IEEE 300 bus system.}
     \label{fig:result_300}
\end{figure}

From now on, we focus on analyzing the impact of the employed weight-bias initialization. Fig.~\ref{fig:weight_histogram} shows the histogram of the relaxed line status values produced by an untrained DA‑DNN under two different initializations. With the proposed manual scheme, almost every value is 0.9999, corresponding to an all‑lines‑closed topology; in contrast, a standard random initializations scatters the outputs across the entire $[0,1]$ range. Since we prepare dataset that are feasible in DC-OPF, the all‑closed topology is always feasible. Thus, the manual initializations guarantees that the very first forward pass can be trained through the embedded DC‑OPF, whereas the random approach leads to infeasible OPF and blocks the learning process.

Fig.~\ref{fig:result_73} and Fig.~\ref{fig:result_300} show the resulting training curves for the IEEE 73‑ and 300‑bus systems. At epoch 0, the embeded optimization problem coincides exactly with the DC‑OPF benchmark, and thus training starts from a feasible dispatch. The generation cost then decreases smoothly without  too much divergence. This behavior cannot be observed when random initial weights is used; in that case, the embedded DC-OPF problem is always infeasible in the initial training, which prevents any learning progress. These results implies that the manual weight–bias initialization is therefore essential, since it secures a feasible starting point to train the network to learn a cost‑reducing switching strategy while satisfying all the network constraints throughout training. 

\section{Conclusion}
In this paper, we proposed a dispatch‑aware deep neural network (DA‑DNN) that consists of line switching layer and embedded differentiable DC‑OPF layer. These two layers enable fully unsupervised training while guaranteeing feasibility at every training step and inference. Also, a customized weight–bias initialization ensures that learning starts from a valid operating point, even on large grids. After training is completed, DA‑DNN outputs a topology–dispatch pair in the same computation time as a baseline DC‑OPF, while achieving the economic benefits of  DC‑OTS . The simulation results on the IEEE 73 and IEEE 300 bus system demonstrated that DA-DNN consistently lowers generation cost relative to DC-OPF. By contratst, commercial solvers for DC-OTS times out on the 300-bus case, which underscores the scalability and real-time sutiability of the proposed method.

For future work, we plan to extend the framework in three directions. First, we will replace the DC-OPF layer with an AC‑OPF layer and incorporate security constraints. Second, we plan to develop a multi‑step switching schedules under forecast uncertainty in load, renewables and contingencies. Lastly, we will incorporates the stability constraints for the applicability of the real-world OTS problem.

\bibliographystyle{IEEEtran}
\bibliography{lgcn_dlr.bib}
\end{document}